\documentclass[draftclsnofoot,a4paper,12pt,onecolumn,doublespace]{IEEEtran}

\usepackage{graphicx}
\usepackage{epsfig}
\usepackage[cmex10]{amsmath}
\usepackage{array}
\usepackage{fixltx2e}

\begin{document}
\title{Tranceiver Design using Linear Precoding in a Multiuser MIMO System with Limited Feedback}

\author{\IEEEauthorblockN{Muhammad Nazmul Islam\IEEEauthorrefmark{0}
 and Raviraj Adve\\ \IEEEauthorrefmark{0}
}
\IEEEauthorblockA{\IEEEauthorrefmark{0}Electrical and Computer Engineering,
University of Toronto, Canada\\
Email:nislam@comm.utoronto.ca, rsadve@comm.utoronto.ca}
}


\maketitle
\begin{abstract}
We investigate quantization and feedback of channel state information in a multiuser (MU) multiple input multiple output (MIMO) system. Each user may receive multiple data streams. Our design minimizes the sum mean squared error (SMSE) while accounting for the imperfections in channel state information (CSI) at the transmitter.
This paper makes three contributions: first, we provide an end-to-end SMSE transceiver design
that incorporates receiver combining, feedback policy and transmit precoder design with channel uncertainty. This 
enables the proposed transceiver to outperform the previously derived limited feedback MU 
linear transceivers. Second, we remove dimensionality constraints on the MIMO system, for the scenario with multiple data streams per user, using a combination of maximum expected signal combining (MESC) and minimum MSE receiver. This makes the feedback of each user independent of the others and the resulting feedback overhead scales linearly with the number of data streams instead of the number of receiving antennas. Finally, we analyze SMSE of the proposed algorithm
at high signal-to-noise ratio (SNR) and large number of transmit antennas. As an aside, we show analytically
why the bit error rate, in the high SNR regime, increases if quantization error is ignored.
\end{abstract}

\begin{IEEEkeywords}
MIMO broadcast channels, Limited feedback of CSI, Quantization error, Ceiling effect, Maximum squared inner product vector quantization, Maximum expected signal combining.
\end{IEEEkeywords}

\section{Introduction}
The advantages of spatial diversity and multiplexing has led to the investigation of multi user (MU) multiple input single output (MISO) and multiple input multiple output (MIMO) wireless communication systems~\cite{Tse}. Spatial diversity can increase system reliability as well as the spectral efficiency of multiuser systems. However, limitations caused by interference and channel fading remain a concern in MU MISO and MU MIMO systems. These can be mitigated by precoding the signals before transmission, in turn requiring channel state information at the transmitter (CSIT). This paper focuses on the linear transceiver design to minimize the sum mean squared error (SMSE) in the downlink of MU MISO~\cite{Schubert:a,Schubert:c} and MIMO systems~\cite{Khachan,Adam}, a single base station (BS) communicating with multiple receivers. 

In a frequency division duplexing (FDD) system, different frequency bands are allocated to the downlink and uplink of a MIMO channel. Therefore, channel information needs to be estimated at the receiver and sent back to the BS after quantization. Recent works suggest that this might be required in a broadband time division duplex (TDD) systems as well~\cite{Haartsen}. In general, providing accurate CSIT and reducing feedback overhead are important considerations in a linear transceiver design. Our work assumes perfect channel estimation at the receiver end with zero delay error-less feedback and focuses on quantizing CSI.

In the available literature, scalar quantization~\cite{Natasha,Khachan:b,Heath,Narula}, vector quantization (VQ)~\cite{Jindal:b,Boccardi:a} and matrix quantization~\cite{Sloane, Jindal:e} have all been used to quantize CSI. It is now well established in the single user, single data stream, case that projecting the MIMO channel to an appropriate vector downlink channel yields better performance than full channel scalar quantization with same feedback overhead~\cite{Heath:b}. This has led to considerable research in VQ, which reduces the feedback overhead by allocating bits in the propoer vector downlink channel. In VQ, to send $ B $ feedback bits as the channel index to the BS, each user needs a codebook with $ 2^B $ code vectors. Grassmanian line packing~\cite{Love:b}, VQ using MSE as the optimality criterion~\cite{Love:a} and random vector quantization (RVQ)~\cite{Jindal:a} have been the most poplular approaches \emph{for the multiuser case}.
 In this paper we investigate VQ, based on the MSIP
criterion~\cite{Rao} as the feedback method. 

In a MU MISO system, users can feed back the channel vectors using VQ. However, in the MIMO case, one needs to combine the receive antennas to convert the MIMO channel to the effective vector downlink MISO channel. Because the receivers cannot cooperate, the quantization scheme of each user is independent of the other. Projecting the MIMO channel to the direction of its maximum eigen vector (MET)~\cite{Boccardi:a} is the optimal solution at low SNR. Jindal~\cite{Jindal:b,Jindal:d} proposed quantization based computing (QBC) which choses the effective vector downlink channel to produce the least quantization error; this is optimal at high SNR.  Trevellato et. al.~\cite{Boccardi:b} proposed maximum expected signal combining (MESC) to maximize the signal to interference plus noise ratio (SINR). Their scheme outperforms both QBC and MET; MESC converges to MET and QBC in the low and high SNR region respectively~\cite{Boccardi:b}. All these schemes discussed so far assume that each user receives a single data stream. However, our intent here is the general case wherein a user, with multiple receive antennas may receive multiple data streams~\cite{Khachan, Adam}. Multiple data streams per user complicates the feedback process, requiring linearly independent information for each stream. In this paper, we extend the MESC algorithm to multiple data stream scenario.

Most of the relevant works in the limited feedback MU literature suffer from dimensionality constraints. With $ M $ transmit antennas, $ N $ total receive antennas and $ L $ data streams in total, either $ M \geq N $~\cite{Jindal:a} or $ N = L $~\cite{Jindal:c}. To the best of our knowledge, only authors in~\cite{Natasha,Khachan:b,Heath} avoid these constraints. However, by using scalar quantization, the feedback overhead in these systems scale linearly with $ 2MN $~\cite{Natasha, Khachan:b} and $ M^2-1 $~\cite{Heath} respectively. Due to the formulation of our MESC receive
combining, the feedback overhead in our proposed system scales only with $ M \times L $ (where $ L $ is the total number of receive data streams). Since, $ L \leq M $ and $ L \leq N $, the proposed transceiver allows significant performance improvements with lower feedback rate.

Previous works~\cite{Love:a} have shown, by simulation, that if the quantization error is ignored, the MSE increases at high SNR. Here we investigate why this is true theoretically.

The overall contributions of this paper are therefore:

1. We provide an end to end SMSE transceiver design that eliminates the dimensionality constraint and tie the feedback overhead to the number of data streams, which is always less than or equal to both the number of transmit and receive antennas.

2. We extend and make the MESC receiver flexible, by allowing multiple data streams per user scenario.

3. We show the flooring effect in terms of SMSE in multiuser broadcast systems. Previous works on this area focused on the
ceiling effect in terms of capacity~\cite{Jindal:a,Jindal:b} and signal to interference plus noise ratio (SINR)~\cite{Love:a}. As an aside, we show why SMSE and BER increases instead of getting flattened out if quantization error is not considered.

The remainder of this paper is organized as follows. Section II describes the system model and reviews transceiver design problem with full channel knowledge. Section III reviews the proposed quantization method and shows the linear precoder design. Section IV illustrates the two step receiver design process and gives the overall algorithm. We analyze our proposed transceiver in Section V. After providing the numerical simulation results in Section VI, we draw our conclusions in section VII.

\emph{Notation}: Lower case, $ x $, denotes scalar while lower case bold face, $ \mathbf{h} $ means column vector. Upper case boldface, $ \mathbf{V} $ denotes matrix whereas uppercase normal font, $ N $, represents constant entry. The superscripts $ (\cdot)^T $ and $ (\cdot)^H $ denote the transpose and conjugate transpose operators respectively. tr [$\cdot$] denotes the trace operator. $ \mathbf{I} $ is reserved for the identity matrix whereas $ \mathbf{1} $ represents the column with all one vector. $ diag(\cdot) $ denotes the diagonal matrix where the diagonal entries contain the bracketed terms. $ ||\cdot||_1 $ denotes the $ L^1 $ norm of the vector. $ E(\cdot) $ denotes statistical expectation.

\section{System Model}

We consider both MU MISO and MU MIMO systems in our design. We at first describe the MU MIMO system model and show that, with our approach, the transceiver design in the MU MIMO system is very similar to that in a MU MISO system.

\subsection{MIMO system model}

Consider a single base station equipped with $ M $ transmit antennas
communicating $K$ independent users. User $k$ has $N_k$ antennas and receives
$L_k$ data streams. Let $L = \sum_k L_k$, $N = \sum_k N_k$. The
$i^\mathrm{th}$ data stream is processed by a unit norm linear
precoding vector $ \mathbf{u}_i $ with the global precoder $
\mathbf{U} = \left [ \mathbf{u}_1, \mathbf{u}_2,...,\mathbf{u}_L \right
] $. Let $ \mathbf{p} = \left [ p_{1}, p_{2}, .., p_{L} \right]^T$ be
the powers allocated to the $ L $ data streams and define the downlink power
matrix $ \mathbf{P} = diag(\mathbf{p}) $. $||\mathbf{p}||_1 \leq
P_{max} $ where $ P_{max} $ is the total available power. The overall data vector
 is $ \mathbf{x} = \left [x_1, ....,x_L \right ]^T = \left[ \mathbf{x}_1^T, \mathbf{x}_2^T, \dots,
\mathbf{x}_K^T\right]^T$. The $N_k \times M$ block fading channel,
$\mathbf{H}_k^H$, between the BS and the user is assumed to be flat.
The global channel matrix is $ \mathbf{H}^H $, with $ \mathbf{H} =
\left [ \mathbf{H}_1,...,\mathbf{H}_k \right] $. User $ k $ receives
\begin{equation}
\mathbf{y}^{DL}_k = \mathbf{H}^H_k \mathbf{U} \mathbf{\sqrt{P}} \mathbf{x} + \mathbf{n}_k,
\end{equation} %
where $ \mathbf{n}_k $ represents the zero mean additive white Gaussian noise at
the receiver with $ E\left[\mathbf{n}\mathbf{n}^H\right] = \sigma^2
\mathbf{I}_{N_k}$. We also assume, $E\left[\mathbf{x}\mathbf{x}^H\right] =
\mathbf{I}_L$. To estimate its own transmitted symbols, from
$\mathbf{y}^{DL}_k $, user $k$ forms 
$$\hat{\mathbf{x}}_k = \mathbf{V}^H_k \mathbf{y}^{DL}_k $$
 Here $ \mathbf{V}_k $ is the $ N_k \times L_k$ decoder vector for user $ k $.	
Figure 1 shows the block diagram of the proposed system in the downlink. 
Let $ \mathbf{V} $ be the $N \times
L$ block diagonal global decoder matrix, $ \mathbf{V} = diag \left (
\mathbf{V}_1, ..., \mathbf{V}_K \right) $. Overall
\begin{eqnarray}
\hat{\mathbf{x}} & = & \mathbf{V}^H\mathbf{H}^H\mathbf{U}\sqrt{\mathbf{P}}\mathbf{x} + \mathbf{V}^H\mathbf{n} \nonumber\\
& = & \mathbf{F}^H \mathbf{U} \sqrt{\mathbf{P}} \mathbf{x} + \mathbf{V}^H \mathbf{n} 
\end{eqnarray}
where, $ \mathbf{n} = \left[ \mathbf{n}_1^T, \mathbf{n}_2^T, \dots,
\mathbf{n}_K^T\right]^T$ and to facilitate our analysis, we define the $M\times L$ matrix
$\mathbf{F} = \mathbf{HV} $ with $ \mathbf{F} = \left [\mathbf{f}_1,
\dots , \mathbf{f}_L \right]$. The vectors $ \mathbf{f}_1 , \dots,
\mathbf{f}_L $ are the effective $ M \times 1 $ MISO channels for the
individual data streams.  

The MSE of the $i^\mathrm{th}$ data stream is given by,
\begin{equation}
    e^{DL}_i = E \left [ \left ( \widehat{x}_i - x_i \right)
                \left ( \widehat{x}_i - x_i \right)^H \right ]. \label{eqn:downlink_mse}
\end{equation}
The SMSE minimization problem is, %
\begin{equation}
 \min_{\mathbf{p},\mathbf{U}} \sum^L_{i=1} e^{DL}_i; \hspace*{0.1in} \mathrm{subject\ to}
    \hspace*{0.1in} \left | \left | \mathrm{p} \right | \right |_1 \leq P_{max}, ||\mathbf{u}_i||=||\mathbf{v}_i||=1 \label{objective}
\end{equation} 
In designing the precoder $\mathbf{U}$, it is computationally efficient to
use a virtual dual uplink~\cite{Khachan}. In this uplink
the transmit powers are $ \mathbf{q} = \left [q_1, .., q_L \right]^T$
for the $L$ data streams, while the matrices $ \mathbf{U} $ \& $ \mathbf{V} $ remain
the same as before. The global virtual uplink power allocation
matrix $\mathbf{Q}$ is defined as, $ \mathbf{Q} = diag(\mathbf{q}) $ where $||\mathbf{q}||_1 \leq
P_{max} $. So the received data in the BS in the virtual uplink is given by,
\begin{equation}
\mathbf{y}^{UL}=\left(\sum^L_{j=1} \mathbf{f}_j \sqrt{q_j} x_j + \mathbf{n} \right) \label{y_uplink}
\end{equation}
Therefore, data stream $ i $ is decoded as,
\begin{equation}
\hat{x}_i^{UL} = \mathbf{u}^H_i \left(\sum^L_{j=1} \mathbf{f}_j \sqrt{q_j} x_j
                                                                    + \mathbf{n} \right) \label{x_hat}
\end{equation}
Figure 2 shows the proposed system model in the virtual uplink. To ensure resolvability, we assume $ L \leq M $ and $ L_k \leq N_k $. From \eqref{y_uplink}, \eqref{x_hat} and 
Fig.~\ref{fig:MUuplink}, it can be seen that our proposed MU MIMO system has become an effective MU single input multiple
output system in the virtual uplink. 

Uplink Downlink duality states that the same MSEs can be achieved in the uplink and the downlink,
with the same matrices $ \mathbf{U} $ and $ \mathbf{V} $ and the same power constraint. A recent result shows that
at the optimal solution, $ \mathbf{p} = \mathbf{q} $~\cite{Adam:b}. 

With perfect channel knowledge, the transmitter iterates between $ \mathbf{V} $ and $ \mathbf{Q} $ and converges to the the optimum solution using a convex optimization problem formulation~\cite{Khachan}. Then the precoder finds the optimal 
$ \mathbf{U} $ using the MMSE solution~\cite{Khachan}. The downlink power allocation is then set equal to $ \mathbf{Q} $~\cite{Adam:b}.

\section{MSIP quantization and linear precoder design}

\subsection{MSIP quantization}

We assume that the receivers have perfect CSI using training. For the
purposes \emph{of quantization only}, the $i^\mathrm{th}$ user choses the quantized codevectors, $ \mathbf{\hat{f}}_i, \dots, \mathbf{\hat{f}}_{L_i} $  that would maximize 
the SINR of its receiving data streams. Chosing the best quantized codevector is described in details in the receiver design section. Each user has a codebook consisting of $ 2^B $ unit norm vectors $
\hat{\mathbf{w}}_1,...,\hat{\mathbf{w}}_{2^B} $. Each user feeds back $B$ bits per data stream
to the BS. The receivers individually normalize and then quantize each
of the $ \mathbf{L}_k $ effective channels using the chordal distance~\cite{Sloane}.
\begin{equation}
\mathbf{\hat{f}}_i = \arg \min_{ \mathbf{w} \in \mathbf{\hat{w}}_1,..,
        \mathbf{\hat{w}}_{2^B} } \sin^2 \left ( \angle \left (\mathbf{f}_i,\mathbf{\hat{w}}
        \right ) \right) \label{quantize}
\end{equation}
The use of chordal distance over the Euclidean distance leads to a higher inner product between the original and quantized
channels~\cite{Rao}. Here, we only quantize the direction of the effective channel and this direction can lie anywhere in the $ M $-dimensional complex unit-norm sphere. Therefore, we generate the quantization codebook as a VQ problem using the MSIP
optimality criterion~\cite{Rao}. Each user at first generates a large set of random unit norm $ M $-dimensional complex  vectors, $ \mathbf{f} $, and finds the quantizer codebook $ C $ to maximize the MSIP,
\begin{equation}
\left(\mathbf{\hat{w}}_1,\cdots,\mathbf{\hat{w}}_{2^B}\right) = \max_{C \left ( \cdot \right)} E \left |<\mathbf{f},C \left( \mathbf{f} \right) > \right|^2  \label{MSIPgeneration} 
\end{equation}

Here, $ \mathbf{\hat{f}} = C \left( \mathbf{f} \right) $ is the quantized effective channel and $ <\cdot,\cdot> $ denotes
the inner product. Details of MSIP VQ codebook generation can be found in~\cite{Rao}.
Overall, we consider the following channel model at the BS for precoder design,
\begin{equation}
\mathbf{f}_i = \mathbf{\widehat{f}}_i + \mathbf{\widetilde{f}}_i \hspace*{0.1in} \mathrm{or}
\hspace*{0.1in} \mathbf{F} = \mathbf{\widehat{F}} + \mathbf{\widetilde{F}}
\label{channel_model}
\end{equation} %
Here, $\mathbf{F} $ comprises $L$ unit-norm effective channel vectors with the
original channel directions. $ \mathbf{\widehat{F}} $ denotes the $L$
quantized feedback unit norm vectors. $\mathbf{\widetilde{F}}$ denotes the error in the quantization.
We assume that the quantization error matrix $ \mathbf{\widetilde{F}} $ has $
L \times M $ independent identically distributed (i.i.d.) elements with
zero mean and a variance of $\frac{\sigma^2_E}{M}$. Here, $\sigma^2_E$ is the quantization error
associated with each quantized vector $ \mathbf{\hat{f}}_i $. We also assume that
$\mathbf{\widetilde{F}}$ is independent of 
$\mathbf{x}$, $\mathbf{n} $ and
$\mathbf{\widehat{F}}$. %

Since $ \sigma^2_E $ depends on receiver combining, the details regarding the expected value of
this term in our proposed algorithm will be clarified in the receiver design and analysis section.

It should be noted that the channel vector of each user automatically takes the form of 
$ \mathbf{\widehat{f}}_i $ in the MU MISO case. Therefore, in this case, we just quantize the normalized
channel using chordal distance and return the corresponding index to the BS.

\subsection{Linear Precoder Design}

The optimum $ \mathbf{U} $ in the SMSE minimization problem of the system model proposed in \eqref{channel_model} 
and \eqref{x_hat} has been solved in the virtual uplink by~\cite{Shenouda}. 
\begin{eqnarray}
& & \mathbf{u}^{MSE}_i = \mathbf{J}^{-1} \mathbf{\widehat{f}}_i \sqrt{q_i} \label{DLPrecoder} \\
& & \mathbf{J} = \mathbf{\widehat{F}} \mathbf{Q} \mathbf{\widehat{F}}^H +
            \sigma^2 \mathbf{I}_M + \frac{\sigma^2_E}{M} \left(q_1 + .. + q_L \right) \mathbf{I}_M  \label{eq:J} 
\end{eqnarray}
So,
\begin{equation}
 e^{UL,MSE}_i = 1 - \sqrt{q_i} \mathbf{\widehat{f}}^H_i \mathbf{J}^{-1}
                                            \mathbf{\widehat{f}}_i \sqrt{q_i} \label{ULSMSE}
 \end{equation}
Therefore, the uplink SMSE is,
\begin{eqnarray}
SMSE^{UL} & = & \sum^L_{i=1}  e^{UL,MSE}_i \nonumber\\ 
& = &  \sum^L_{i=1} 1 - \sum^L_{i=1} \sqrt{q_i}
            \widehat{\mathbf{f}}^H_i \mathbf{J}^{-1} \widehat{\mathbf{f}}_i \sqrt{q_i} \nonumber\\
    & = &  L - tr \left [ \mathbf{\widehat{F}} \mathbf{Q} \mathbf{\widehat{F}}^H \left ( \mathbf{\widehat{F}} \mathbf{Q} \mathbf{\widehat{F}}^H + \left (\sigma^2 +  \frac{\sigma^2_E \sum^L_{i=1} q_i}{M}  \right) \mathbf{I}_M \right )^{-1} \right ] \nonumber \\
    & = &  L-M+\left (\sigma^2 +\frac{\sigma^2_E \sum^L_{k=1} q_k}{M}  \right)
                        tr \left [ \mathbf{J}^{-1} \right]  \label{eq:SMSE_main}
\end{eqnarray}   
As $\mathbf{\widehat{F}}$ is fixed, the SMSE expression is a function of
uplink power allocation $\mathbf{Q} $.

\emph{Proposition 1 :} The optimization problem for power allocation,
\begin{equation}
\mathbf{Q}^{opt} = \min_{\mathbf{Q}}
    \left(\sigma^2+\frac{q_1+...+q_L}{M}\sigma^2_E\right)tr(\mathbf{J}^{-1}) \label{eq:ULPowerAlloc}
\end{equation}
subject to $ tr [\mathbf{Q}] \leq P_{max} , q_k \geq 0 $ for all $k$ is
convex in $\mathbf{Q}$.

\noindent \emph{Proof}: Ding~\cite{Ding} shows that SMSE remains a nonincreasing 
function of SNR if channel uncertainty is equal and all available power
is used because $ tr(\mathbf{Q}) = \sum_{i=1}^L{q_i} = P_{max} $ makes the term 
within the brackets a constant.$\mathbf{J} $ is a positive definite matrix and therefore, the optimization
problem is convex in $ \mathbf{J} $~\cite{Boyd}. Since $ \mathbf{J} $ is linear
in $\mathbf{Q}$, it can be readily proved that the problem is convex in
$ \mathbf{Q} $.

The power allocation problem is therefore convex given $ \mathbf{\widehat{F}} $. In the next section we discuss the remaining
problem, the solution for $ \mathbf{V} $ (equivalently $ \mathbf{F} $). This section represents the 
core contribution of the paper.

\section{Receiver Design}

We propose a two step receiver design. For the
purposes \emph{of quantization only}, each user uses a MESC receiver and 
choses the quantized codevectors that would maximize the SINR of their data streams. 
However, the users implement MMSE receivers while receiving the actual data. This is unlike the single MMSE solution
in~\cite{Khachan, Schubert:a},  but allows for the \emph{channel
feedback to be independent of the other users' actions}. 

\subsection{Receive combining with MESC}

Before going into the analysis, let us clarify the relation between $ \mathbf{F} $ and $ \mathbf{HV} $ 
that will be used interchangably in this section. We assume $ \mathbf{f}_i = \mathbf{H}_k \mathbf{v}_i $ where  $ \mathbf{H}_k $ is the channel of the $ k^\mathrm{th}$
user receiving the $ i^\mathrm{th}$ data stream and $ \mathbf{v}_i $ is the decoding vector used for the $ i^\mathrm{th}$ stream; $ \mathbf{u}_i $ and $ \mathbf{u}_j $ are the precoding vectors of the $ i^\mathrm{th}$ and $ j^\mathrm{th}$ data. 

Now, using our quantization policy in \eqref{quantize}, we define the
quantization angle $ \theta_i \in [0,\frac{\pi}{2}] $ as,
\begin{equation}
\cos \theta_i = \left | \mathbf{\overline{f}}_i^H \mathbf{\hat{f}}_i \right |  \label{quantization_angle}
\end{equation}
Here, $  \mathbf{\overline{f}}_i $ represents the unit norm effective vector downlink channel i.e.
$ \mathbf{\overline{f}} = \frac{\mathbf{f}}{||\mathbf{f}||} $. Here $ \mathbf{f} $ is the effective MISO channel for the
data stream.

Since, the receivers know exactly the quantization angle, we can use this information
to improve the expected SINR. As in~\cite{Boccardi:b}, 
define the quantization error as,
\begin{equation}
\mathbf{\widetilde{f}}_i = \mathbf{\overline{f}}_i - \left(\mathbf{\widehat{f}}_i^H \mathbf{\overline{f}}_i \right) \mathbf{\widehat{f}}_i \label{eq:quantization_error_receiver}
\end{equation}
It can be easily verified that $ ||\mathbf{\widetilde{f}}_i||^2 = \sin^2 \theta_i $. Now the SINR in the downlink for the $i^\mathrm{th}$ stream is,
\begin{equation}
\mathrm{SINR}_i^{DL}  =  \frac{\frac{P}{L} \left |\mathbf{f}_i^H \mathbf{u}_i \right |^2}{\sigma^2 + \sum_{j \in L, j \not= i} \frac{P}{L} \left |\mathbf{f}_i^H \mathbf{u}_j  \right |^2}    \label{eq:exprected_SINR_one} 
\end{equation}
In \eqref{eq:exprected_SINR_one} equal power allocation was assumed to simpify the receiver combining analysis. Here, $ \mathbf{u}_i $ and $ \mathbf{u}_j $ follows the form in \eqref{DLPrecoder}. Now using \eqref{DLPrecoder} and the matrix inversion lemma~\cite{Strang}, 
\begin{eqnarray}
\mathbf{u}_i & = & \left ( \left (\sigma^2 + \frac{\sigma^2_E}{M} P_{max} \right) \mathbf{I} + \mathbf{\widehat{F}} \mathbf{Q} \mathbf{\widehat{F}}^H \right)^{-1} \mathbf{\hat{f}}_i \sqrt{q_i}  \nonumber\\
& = & \frac{1}{\left (\sigma^2 + \frac{\sigma^2_E}{M} P_{max} \right)} \mathbf{\hat{f}}_i \sqrt{q_i} - \frac{1}{\left (\sigma^2 + \frac{\sigma^2_E}{M} P_{max} \right)^2}  \times  \nonumber\\
& & \mathbf{\widehat{F}} \left (\mathbf{Q}^{-1}+\frac{1}{\sigma^2+\frac{\sigma^2_E}{M}P_{max}} \mathbf{\widehat{F}}^H \mathbf{\widehat{F}} \right)^{-1}  \mathbf{\widehat{F}}^H \mathbf{\widehat{f}}_i \sqrt{q_i} \label{eq:woodbury_one}
\end{eqnarray}
Here, $ ||\mathbf{u}_i || = 1 $. Since the users do not cooperate in our scheme, the quantization and feedback methods implemented by the users need to be independent of each other. Since the effective MISO channels of different users are statistically independent of each other, we assume different user's quantized channels to be mutually orthogonal i.e. 
$ \mathbf{\hat{f}}_i^H \mathbf{\hat{f}}_j = 0$ where $ i $ and $ j $ indicate data streams that are being received by two different users. Following this assumption, it can be easily verified from \eqref{eq:woodbury_one} that,$ \mathbf{\hat{f}}_i^H \mathbf{u}_j = 0 $.

Since each user knows the inner product of different code vectors in its codebook, the assumption of orthogonality is not valid for two different streams of the same user. Therefore, in our proposed algorithm, each user uses its known codevectors, i.e. the effective channels of its data streams, as a set of column vectors $ \mathbf{\widehat{f}} $ in the $ \mathbf{\widehat{F}} $ matrix and assumes that the vector downlink channels for all other users' stream are mutually orthogonal to its own channels. We also assume that noise variance, signal power, quantization error assumption in the BS and total number of data streams sent by the BS are known to each of the users. Therefore due to the construction of \eqref{eq:woodbury_one}, each user can find the expected value of $ \mathbf{f}_i^H \mathbf{u}_i $ and $ ||\mathbf{u}_i || $ even without co-operating with other users. 
Therefore, by seperating the intra user streams from inter-user streams, \eqref{eq:exprected_SINR_one} takes the following form,
\begin{equation}
SINR_i^{DL} =  \frac {\frac{P}{L} |\mathbf{v}_i^H \mathbf{H}_k^H \mathbf{u}_i |^2}{ \sigma^2 + \sum_{j \in L_k,j \not= i} \frac{P}{L} | \mathbf{v}_i^H \mathbf{H}_k^H \mathbf{u}_j |^2 + \sum_{j \in L,j \not\in L_k} \frac{P}{L} | \mathbf{f}_i^H \mathbf{u}_j |^2} \label{eq:SINR2} 
\end{equation}
Now,
\begin{eqnarray}
\lefteqn{\sum_{j \in L, j\not\in L_k} | \mathbf{f}_i^H \mathbf{u}_j |^2} \nonumber\\
& = & ||\mathbf{f}_i||^2  \sum_{j \in L, j\not\in L_k}  \left|\left| \left(\mathbf{\widehat{f}}_i^H \mathbf{\overline{f}}_i \right) \mathbf{\widehat{f}}_i^H \mathbf{u}_j  + \mathbf{\widetilde{f}}_i^H \mathbf{u}_j  \right|\right|^2 \label{eq:interference1} \\ 
& = & ||\mathbf{f}_i||^2  \sum_{j \in L, j\not\in L_k} \left |\left| \mathbf{\widetilde{f}}_i^H \mathbf{u}_j \right |\right|^2 \label{eq:interference2} \\
& = & ||\mathbf{f}_i||^2 ||\mathbf{\widetilde{f}}_i||^2 \sum_{j \in L, j\not\in L_k} \left |\left| \mathbf{\overline{\widetilde{f}}}_i^H \mathbf{u}_j \right |\right|^2 \label{eq:interference3} \\
& = & ||\mathbf{f}_i||^2 \sin^2 \theta_i \frac{L-L_k}{M-1} \label{eq:interference4} \\
& = & \frac{L-L_k}{M-1} \left(||\mathbf{f}_i||^2 - \left (\mathbf{f}_i^H \mathbf{\widehat{f}}_i \right) \left (\mathbf{\widehat{f}}_i^H \mathbf{f}_i \right) \right) \label{eq:interference5} \\
& = & \frac{L-L_k}{M-1} \mathbf{v}_i^H \left (\mathbf{H}_k^H \left (\mathbf{I}-\mathbf{\hat{f}}_i \mathbf{\hat{f}}_i^H \right) \mathbf{H}_k \right) \mathbf{v}_i  \label{eq:interference6}
\end{eqnarray}
\eqref{eq:interference1} is obtained by taking out the norm of $ \mathbf{f}_i $ and using \eqref{eq:quantization_error_receiver}. \eqref{eq:interference2} follows since $ \mathbf{\widehat{f}}_i^H \mathbf{u}_j = 0 $ for mutually orthogonal reported channels from different users. \eqref{eq:interference3} was obtained by assuming
$ \mathbf{\overline{\widetilde{f}}}_i = \frac{\mathbf{\widetilde{f}}_i}{||\mathbf{\widetilde{f}}_i||} $. \eqref{eq:interference4} was derived using the analysis of \cite{Boccardi:b}. In the presence of large number of codevectors, $ \theta_i $ is very small which leads to $ \mathbf{\widetilde{f}}_i^H \mathbf{\widehat{f}}_i \approx 0 $. Therefore, the unit vectors $ \mathbf{\overline{\widetilde{f}}}_i $ and $ \mathbf{u}_j $ are both identically distributed in the $ M-1 $ dimensional plane orthogonal to $ \mathbf{\widehat{f}}_i $. Therefore, $ || \mathbf{\overline{\widetilde{f}}}_i^H \mathbf{u}_j ||^2 $ follows a beta distribution with parameter $ (1,M-1) $ and has expected value $ \frac{1}{M-1} $~\cite{Boccardi:b}. The factor of $ L-L_k $ arises since the $ k^\mathrm{th}$ user is receiving $ L_k $ data streams and therefore $ L - L_k $ data streams are mutually orthogonal to the  $ i^\mathrm{th}$ data stream. \eqref{eq:interference5} was obtained using the quantization angle definition from of \eqref{quantization_angle}.  In \eqref{eq:interference6}, we again use $ \mathbf{f}_i = \mathbf{H}_k \mathbf{v}_i $.

Using the results of \eqref{eq:interference6} in \eqref{eq:SINR2} and defining
\begin{equation}
\mathbf{B}_i = \frac{P}{L} \mathbf{H}_k^H \left ( \sum_{j \in L_k,j \not= i} \mathbf{u}_j \mathbf{u}_j^H + \frac{L-L_k}{M-1} \left ( \mathbf{I} - \mathbf{f}_i \mathbf{f}_i^H \right) \right) \mathbf{H}_k  \label{eq:B_i}
\end{equation}
\eqref{eq:SINR2} takes the following form,
\begin{equation}
SINR_i^{DL} = \frac{\mathbf{v}_i^H \frac{P}{L} \mathbf{H}_k^H \mathbf{u}_i \mathbf{u}_i^H \mathbf{H}_k \mathbf{v}_i}{\sigma^2 + \mathbf{v}_i^H \mathbf{B}_i \mathbf{v}_i}
\label{eq:expected_SINR_four}  
\end{equation}
Due to the structure of $ \mathbf{u}_i $ and $ \mathbf{B}_i $, $ SINR_i^{DL} $ in \eqref{eq:expected_SINR_four} is a function 
of $ \mathbf{{v}}_i $  and $ \mathbf{\hat{f}}_j  \forall j \in L_k $. Each of these $ \mathbf{\hat{f}}_j $ vectors are in the codebook $ C $ which consists of $ \mathbf{\hat{w}}_1,\cdots,\mathbf{\hat{w}}_{2^B} $ codevectors. Therefore, the linear decoding vector $ \mathbf{v}_j $ and $ \mathbf{\hat{w}}_j \forall j \in L_k $ should be chosen jointly as,
\begin{equation}
\left(\mathbf{\hat{w}}_j, \mathbf{v}_j \right) \forall j \in L_k = \max_{||\mathbf{v}_j||=1, \mathbf{\hat{w}}_j \in W} \sum_{j=1}^{L_k} SINR_j^{DL}  \label{eq:joint_SINR_MESC}
\end{equation}
Here, $ \mathbf{v}_j $ is a complex $ N_k $ dimensional vector. Joint optimization for all the data streams of a particular user in \eqref{eq:joint_SINR_MESC} will lead to a computational complexity proportional to $ \left(2^B \right)^{L_k} $. One sub-optimal solution to reduce complexity is to find the optimum decoding vector and quantized channel one data stream at a time. This simplified algorithm is given below:

1. First, assume that intra-user streams are orthogonal to find the vector downlink channel of the first stream. Therefore, maximizing \eqref{eq:expected_SINR_four} becomes an optimization problem of $ \mathbf{\hat{w}}_{L_{k_1}} $ and $ \mathbf{v}_{L_{k_1}} $ where $ L_{k_1} $ denotes the first data stream of the $k^\mathrm{th}$ user. So,
\begin{equation}
\mathbf{B}_{L_{k_1}} = \left (\frac{P}{L} \mathbf{H}_k^H \left ( \frac{L-1}{M-1} \left ( \mathbf{I} - \mathbf{\hat{w}}_{L_{k_1}} \mathbf{\hat{w}}_{L_{k_1}}^H \right) \right) \mathbf{H}_k \right )  \label{eq:B_1}
\end{equation}
\begin{equation}
SINR_{L_{k_1}}^{DL} = \frac { \mathbf{v}_{L_{k_1}}^H \left ( \frac{P}{L} \mathbf{H}_k^H \mathbf{u}_{L_{k_1}} \mathbf{u}_{L_{k_1}}^H \mathbf{H}_k \right) \mathbf{v}_{L_{k_1}} } { \sigma^2 + \mathbf{v}_{L_{k_1}}^H \mathbf{B}_{L_{k_1}} \mathbf{v}_{L_{k_1}}}  \label{eq:SINR_MESC_1st_stream}
\end{equation}
\begin{equation}
\left(\mathbf{\hat{w}}_{L_{k_1}}, \mathbf{v}_{L_{k_1}} \right)  = \max_{\left(||\mathbf{v}_{L_{k_1}}||=1, \mathbf{\hat{w}}_{L_{k_1}} \in W \right)} SINR_{L_{k_1}}^{DL}  \label{eq:chose_MESC_1st_stream}
\end{equation}

2. Once the quantized channel for the 1st stream is chosen, the user assumes this to be a nonorthogonal channel for the second stream's vector downlink channel. However, vector downlink channels for the other streams of the same user are still considered to be orthogonal to both first and second stream's channel. Thus maximizing \eqref{eq:expected_SINR_four} again becomes an optimization problem with variable $ \mathbf{v}_{L_{k_2}} $ and $ \mathbf{\hat{w}}_{L_{k_2}} $ for the present data stream where $ L_{k_2} $ denotes the second stream of the $k^\mathrm{th}$ user. So,
\begin{equation}
\mathbf{B}_{L_{k_2}} = \left (\frac{P}{L} \mathbf{H}_k^H \left ( \mathbf{u}_{L_{k_1}} \mathbf{u}_{L_{k_1}}^H + \frac{L-2}{M-1} \left ( \mathbf{I} - \mathbf{\hat{w}}_{L_{k_2}} \mathbf{\hat{w}}_{L_{k_2}}^H \right) \right) \mathbf{H}_k \right )   \label{eq:B_2}
\end{equation}
\begin{equation}
SINR_{L_{k_2}}^{DL} = \frac { \mathbf{v}_{L_{k_2}}^H \left ( \frac{P}{L} \mathbf{H}_k^H \mathbf{u}_{L_{k_2}} \mathbf{u}_{L_{k_2}}^H \mathbf{H}_k \right) \mathbf{v}_{L_{k_2}} } { \sigma^2 + \mathbf{v}_{L_{k_2}}^H \mathbf{B}_{L_{k_2}} \mathbf{v}_{L_{k_2}}}  \label{eq:SINR_MESC_2nd_stream}
\end{equation}
\begin{equation}
\left(\mathbf{\hat{w}}_{L_{k_2}}, \mathbf{v}_{L_{k_2}} \right)  = \max_{\left(||\mathbf{v}_{L_{k_2}}||=1, \mathbf{\hat{w}}_{L_{k_2}} \in W \right)} SINR_{L_{k_2}}^{DL}  \label{eq:chose_MESC_2nd_stream}
\end{equation}
  \newline
3. For the 3rd data stream of the $k^\mathrm{th}$ user,
\begin{equation}
\mathbf{B}_{L_{k_3}} = \left (\frac{P}{L} \mathbf{H}_k^H \left ( \mathbf{u}_{L_{k_1}} \mathbf{u}_{L_{k_1}}^H + \mathbf{u}_{L_{k_2}} \mathbf{u}_{L_{k_2}}^H + \frac{L-3}{M-1} \left ( \mathbf{I} - \mathbf{\hat{w}}_{L_{k_3}} \mathbf{\hat{w}}_{L_{k_3}}^H \right) \right) \mathbf{H}_k \right )   \label{eq:B_2}
\end{equation}
The other equations will take the similar form of the ones mentioned in the previous two data stream's cases. The same policy will be continued upto the last stream of the $ k^\mathrm{th} $ user. Note that as we increase the assumption of the number of data streams in the reported nonorthogonal channels part, the number of components in the summation term $ \sum \mathbf{u}_j \mathbf{u}_j^H $ increases and  $ \frac{L-L_k}{M-1} $ decreases. This follows the reasonings explained in the derivation of \eqref{eq:interference1} to \eqref{eq:interference6}.

With this algorithm, the SINR expression for a particular data stream remains a function of only its decoding vector and its quantized channel. This leads to a computational complexity of $ L_k \times 2^B $ in finding the channels of $ L_k $ data streams. Now \eqref{eq:B_i} and \eqref{eq:expected_SINR_four} can be thought as a general form of all the data stream's SINR expressions. In \eqref{eq:B_i} and \eqref{eq:expected_SINR_four}, both $ \mathbf{f}_i $ and $ \mathbf{u} $ depend on chosen codevector $ \mathbf{\hat{w}}_i $. For any particular $ \mathbf{\hat{w}}_i $, the linear decoding vector that maximizes \eqref{eq:expected_SINR_four} can be obtained by the MMSE detector, $ \mathbf{v}_i = \left (\sigma^2 \mathbf{I} + \mathbf{B}_i \right)^{-1} \sqrt{\frac{P}{L}} \mathbf{H}_k^H \mathbf{u}_i $~\cite{Boccardi:b}. Then,
\begin{equation}
SINR_i^{DL} = \frac{P}{L} \mathbf{u}_i^H \mathbf{H}_k \left (\sigma^2 \mathbf{I} + \mathbf{B}_i \right)^{-1} \mathbf{H}_k^H \mathbf{u}_i  \label{eq:expected_SINR_final}
\end{equation}
The user finds the value of $ SINR_i^{DL} $ for every $ \mathbf{\hat{w}}_i $ using \eqref{eq:expected_SINR_final} and choses the $ \mathbf{\hat{w}}_i $, as quantized channel $ \mathbf{\hat{f}}_i $, that maximizes $ SINR_i^{DL} $.

It is worth emphasizing that, to our knowledge, this is the first scheme that considers signal power, inter-user and intra-user
interference while accounting for multiple data stream per user.

\subsection{Receiver Design for data processing}

As mentioned earlier, MESC is done for
quantization purposes only. The base station determines $\mathbf{p}$
and $\mathbf{U}$ based on the quantized $\mathbf{\widehat{F}}$. However, for mutually nonorthogonal reported channels 
and a finite number of users, using MMSE receivers for data processing provide better results than MESC receivers~\cite{Boccardi:b}. Therefore,
\begin{equation}
\mathbf{v}_i = \left (\mathbf{H}^H_k \mathbf{U} \mathbf{P} \mathbf{U}^H \mathbf{H}_k +
                        \sigma^2 \mathbf{I} \right)^{-1} \mathbf{H}^H_k \mathbf{u}_i \sqrt{p_i}, \label{eq:mmse_decoder1}
\end{equation}   
which can be normalized to make $ ||\mathbf{v}_i || = 1$. $ \mathbf{H}^H_k $ is the MIMO channel of the $ k^\mathrm{th}$ user receiving the $ i^\mathrm{th}$ data stream. $ \mathbf{u}_i $ and $ p_i $ respresent the designed precoder and allocated power for the $ i^\mathrm{th}$ stream. Note that the MMSE receiver cannot be implemented at the time of channel quantization since the precoder matrix $ \mathbf{U} $ was not designed at that time.

The implementation of the decoder mentioned in \eqref{eq:mmse_decoder1} requires infinite dedicated symbol training. Therefore, from a practical point of view, the BS either sends a finite number of dedicated symbols~\cite{Caire} or uses limited feedforward~\cite{Heath:c} to convey the post-processing information to the receivers. However, in our simulations, we
restrict ourselves to the case where the users can estimate the effective channels of their data streams.
 
\subsection{Overall Algorithm}
 
Using the developments in section (III, IV-A and IV-B), the steps of the proposed overall algorithm for SMSE minimization in the MU system are:

1. Send common pilots to the users in the system so that each user can estimate its own channel. \\
2. Each user generates a separate codebook of $ 2^B$ unit norm vectors using
MSIP VQ in off-line. In the MU MIMO case, each user converts its estimated MIMO channel to effective MISO
channels using the MESC algorithm proposed in section IV-A and sends the codebook indexes of the effective channels to the BS. In a MU MISO system, each user 
quantizes its own channel and the BS assumes $ \mathbf{V} = \mathbf{I}$.  \\
3. Virtual uplink power allocation: \\
\hspace*{0.2in} $ \mathbf{Q}^{opt} = \min_\mathbf{Q}
\left (\sigma^2 +\frac{\sigma^2_E P_{max}}{M}  \right) tr(\mathbf{J}^{-1}) $, 
is convex in $ \mathbf{Q} $. Here, $ \mathbf{J} $ follows \eqref{eq:J}. \\
4. Uplink beamforming: $ \mathbf{u}_i = \mathbf{J}^{-1}
\hat{\mathbf{f}}_i \sqrt{q_i} $, $ || \mathbf{u}_i || = 1 $ \\
5. Downlink power allocation $ \mathbf{P} =  \mathbf{Q} $. \\
6. Send dedicated pilot symbols for each of the data streams. Thereafter, implement the MMSE downlink decoders using \eqref{eq:mmse_decoder1}. $ || \mathbf{v}_i || = 1 $ for the $ i^\mathrm{th}$ data stream.

The algorithm above results in a precoder $\mathbf{U}$, decoder
$\mathbf{V}$ and power allocation, $\mathbf{p}$. Note that the solution
is sub-optimal because $\mathbf{U}$ and $\mathbf{p}$ are designed using
MESC, not MMSE.

\section{Analysis \& Discussion}

\subsection{Relation to the existing algorithms}

As the proposed receive combining technique maximizes the expected SINR of the data streams at 
the user end, it is equivalent to the MESC algorithm in the case of one data stream per user of ~\cite{Boccardi:b} which was designed for the ZF precoder. To illustrate this, let $ L_k = 1 $. Since intra user interference is not present, all the quantized effective channels in $ \mathbf{\widehat{F}} $ are assumed to be mutually orthogonal.
Using this in \eqref{eq:woodbury_one} we get,
\begin{eqnarray}
\mathbf{u}_i & = & \frac{1}{\sigma^2+\sigma^2_E P_{max}}\mathbf{\hat{f}}_i \sqrt{q_i} - \frac{1}{\left(\sigma^2+\sigma^2_E P_{max}\right)^2} \mathbf{\hat{F}} \times  \nonumber\\
& & \left ( \mathbf{Q}^{-1} + \frac{1}{\sigma^2+\sigma^2_E P_{max} \mathbf{I}} \right)^{-1} [1,0,\cdots,0]^T \sqrt{q_i} \nonumber\\
& = & c \times \mathbf{\hat{f}}_i,  \label{eq:MESC_one_stream}
\end{eqnarray}
Where $ c $ is some constant. \eqref{eq:MESC_one_stream} follows since $ \left ( \mathbf{Q}^{-1} + \frac{1}{\sigma^2+\sigma^2_E P_{max}} \mathbf{I} \right)^{-1} $ is a diagonal matrix. Since $ ||\mathbf{u}_i || = 1 $, $ \mathbf{u}_i = \mathbf{\hat{f}}_i $ in this scenario. Using this in \eqref{eq:SINR2} we find,
\begin{equation}
SINR_i^{DL} = \frac { \mathbf{v}_i^H \left ( \frac{P}{L} \mathbf{H}_k^H \mathbf{\hat{w}}_i \mathbf{\hat{w}}_i^H \mathbf{H}_k \right) \mathbf{v}_i } { \sigma^2 + \mathbf{v}_i^H \left (\frac{P}{L} \mathbf{H}_k^H \left ( \frac{L-1}{M-1} \left ( \mathbf{I} - \mathbf{\hat{w}}_i \mathbf{\hat{w}}_i^H \right) \right) \mathbf{H}_k \right ) \mathbf{v}_i } 
\label{eq:expected_SINR_one_stream}
\end{equation}
This is the exact same expression obtained in \cite{Boccardi:b} as the MESC combiner with noise variance $ \sigma^2 = 1 $.   \cite{Boccardi:b} has shown that this algorithm takes the form of MET combining at low SNR and QBC at high SNR. Thus MESC combining of~\cite{Boccardi:b} considers signal power and inter user intereference while chosing the code vector. Since we are considering multiple data streams to each user, our proposed SINR expression in \eqref{eq:SINR2} considers signal power, inter user and intra user interference altogether. Thus our proposed algorithm is a generalized form for MESC combining with multiple data streams.

\subsection{Quantization error analysis}

Due to the structure of the receive combining, the quantization error in the quantized feedback effective MISO channel varies from low to high SNR. Thus, the variance of $\mathbf{\tilde{f}}_i $ varies, too. In the following, we give a brief explanation of the quantization error variance in the high and low SNR scenario. \newline

\emph{Quantization Error at Low SNR}:

In the low SNR region, we can asseme, 
$ \sigma^2 \gg \sum_{j \in L, j \not= i} \frac{P}{L} \left |\mathbf{f}_i^H \mathbf{u}_j  \right |^2 $
in \eqref{eq:exprected_SINR_one}. Therefore, the proposed scheme leads to maximizing signal power. 
Thus, the quantization problem can be formulated as finding the decoding vector that 
would maximize the signal power and then finding the quantized code vector that
is closest to the newly formed vector downlink MISO channel.

Due to the formulation of the MSIP approach, the error variance of quantization error, $ \sigma^2_E $, is measured in terms
of the angle spread between the original and quantized vectors. In~\cite{Rao}, the quantization error of $ \mathbf{\tilde{f}} $ was given the following form,
\begin{equation}
\sigma^2_E = E \left [ \sin^2 \left(\angle
\left(\mathbf{f}_i,\mathbf{\widehat{f}}_i \right) \right) \right ] \leq 2^{\frac{-B}{M-1}} \label{eq:MSIPerror}
\end{equation}
Since we are only quantizing the direction, not magnitude, this error variance
denotes the angular spread of the quantized effective MISO channel. 

\emph{Quantization error at high SNR}:

In Section V-A, we have shown our proposed algorithm is equivalent to QBC at high SNR for one data stream per user. Section VI will show the simulation of the convergence of this algorithm to QBC for multiple data streams per user. Therefore, we analyze the high SNR quantization error of  our receiving combining scheme using the concepts of QBC.  

When each user receives one data stream, QBC choses the codevector with the least quantization error and thus converts a MIMO channel into an effective MISO channel~\cite{Jindal:b}. The quantization error in this case is upper bounded by $ 2^{\frac{-B}{M-N_k}} $~\cite{Jindal:b}. Using the same notion, for a multiple data stream per user scenario, the effective MISO channel of the $i^\mathrm{th}$ stream of a particular user can be chosen to generate the $i^\mathrm{th}$ least quantization error with respect to its original MIMO channel. The expected quantization error of the $i^\mathrm{th}$ data stream (in terms of error tolerance) of the $k^\mathrm{th}$ user in this method satisfies~\cite{Jindal:b,Gupta}, 
\begin{equation}
\sigma^2_E \leq i \times 2^{\frac{-B}{M-N_k}}    \label{eq:quanterror_QBC}
\end{equation}
Note that the quantization method described in the previous passage can lead to intra-user interferrence due to the correlation of two codevectors of a particular codebook. Our proposed algorithm avoids this scenario by incorporating intra user interference effect in receiver combining. However, the codevectors chosen for two different streams of a user vary with time and become statistically independent with each other over long term channel realizations. Therefore, we hypothesize that the quantization error of our algorithm matches with the one given in \eqref{eq:quanterror_QBC} at high SNR.

The proposed receive combining scheme incorporates both an increase in signal power and reduction in (intra and inter user) interference. The trade-off between these two depends on the SNR. Due to the adaptive nature of this method, the expected quantization errors for intermediate SNR cases are very hard to derive. In our simulations we assumed the quantization error to take the form of \eqref{eq:MSIPerror} at low SNR (0 dB) and changed this value linearly with transmitted power so that it converged to the form of \eqref{eq:quanterror_QBC} at high SNR (30 dB). Investigation regarding the exact value of the expected quantization error at the intermediate SNR remains an open area of future research.

In summary, the quantization error of the proposed algorithm ranges between $ 2^{\frac{-B}{M-1}} $ and $ i \times 2^{\frac{-B}{M-N_k}} $. Note that, in most of the cases, both these error variances are lower than the errors in VQ MSE (which quantizes both magnitude and direction) with $ \sigma^2_E \geq 2^{\frac{-B}{M}} $~\cite{Love:a}. Therefore, the proposed algorithm quantizes the channel directions more precisely than the previously proposed VQ MSE feedback policy.

\subsection{SMSE Analysis}

In the absence of quantization error, SMSE of the traditional precoder
\cite{Khachan} (where quantization error is not considered) is
\begin{eqnarray}
& & SMSE = L-M+\sigma^2 tr \left [ \left (\mathbf{F} \mathbf{Q} \mathbf{F}^H +\sigma^2 \mathbf{I}_M \right)^{-1} \right]  \nonumber\\
& & = L-M+ tr \left [ \left ( \frac{P_{max}}{L \sigma^2} \mathbf{F} \mathbf{F}^H  + \mathbf{I}_M \right)^{-1} \right] \label{eq:IdealMSE2}
\end{eqnarray}
In \eqref{eq:IdealMSE2}, we assumed $ \mathbf{Q} = \frac{P_{max}}{L} \mathbf{I}_L $ i.e. equal power 
allocation for simplicity of the analysis. At very high SNR, the SMSE approaches zero in \eqref{eq:IdealMSE2} as $
tr \left ( \frac{P_{max}}{L \sigma^2} \mathbf{F} \mathbf{F}^H  + \mathbf{I}_M \right)^{-1} $ is a decreasing function of SNR.
However, with quantization error, if the original
precoder~\cite{Khachan} is used,
\begin{equation}
SMSE = \sum^L_{i=1} \left( 1 - q_i \mathbf{\widehat{f}}^H_i \mathbf{J}^{-1}
            \mathbf{\widehat{f}}_i + \frac{\sigma^2_E}{M} P_{max} q_i \mathbf{\widehat{f}}^H_i
            \mathbf{J}^{-2} \mathbf{\widehat{f}}_i \right)
\end{equation}
where $\mathbf{J} = \mathbf{\widehat{F}} \mathbf{Q}
\mathbf{\widehat{F}}^H +\sigma^2 \mathbf{I}_M$. Both $ q_i
\mathbf{\widehat{f}}^H_i \mathbf{J}^{-1} \mathbf{\widehat{f}}_i$ and
$\frac{\sigma^2_E}{M} P_{max} q_i \mathbf{\widehat{f}}^H_i
\mathbf{J}^{-2} \mathbf{\widehat{f}}_i $ increase with SNR. Since the
former term is a linear over affine function and the latter is a
quadratic over quadratic function of $ P_{max} $, at high SNR the latter term
dominates and SMSE increases with SNR, which explains the results
of~\cite{Love:a}.

In our proposed algorithm, 
\begin{eqnarray}
& & \hspace{-7mm} SMSE = L-M+ \left(\sigma^2 + \frac{\sigma^2_E}{M} P_{max} \right) \left [ \left (\mathbf{F} \mathbf{Q} \mathbf{F}^H + \left(\sigma^2 + \frac{\sigma^2_E}{M} P_{max} \right) \mathbf{I}_M \right)^{-1} \right] \label{eq:proposedMSE1} \\
& & \hspace{-7mm} = L - M + tr  \left ( \frac{P_{max}}{L \left (\sigma^2 + \frac{\sigma^2_E}{M} P_{max} \right)} \mathbf{F} \mathbf{F}^H + \mathbf{I}_M \right)^{-1}  \label{eq:proposedMSE2}  
\end{eqnarray}
In \eqref{eq:proposedMSE2}, we again assumed equal power allocation for analysis.
$ \frac{P_{max}}{L \left (\sigma^2 + \frac{\sigma^2_E}{M} P_{max} \right)} $ is a nonincreasing
function of $ P_{max} $. Thus the proposed precoder makes sure that
SMSE does not increase with SNR at high SNR region. Fig~\ref{fig:SMSE_analysis}
illustrates all these effects. Since, the increase in SMSE is most apparent in MU-MISO systems due 
to their lack of diversity, the simulations are done in a MU-MISO system with
independent channel realizations where $ M = 5 $, $ L_k = 1$ $\forall k $ and $ B =
10 $ bits per data stream. The proposed algorithm clearly stabilizes the SMSE at
high SNR. 

Note that, at very high SNR, $ \frac{P_{max}}{L \left (\sigma^2 + \frac{\sigma^2_E}{M} P_{max} \right)} $ will become constant and make SMSE saturated. This leads to the following result. \emph{For a fixed quantization
error, the SMSE of a multiuser system is lower bounded by a fixed value which does not depend 
on SNR}. We call this the \emph{flooring effect} of multiuser broadcast systems. This is similar to the ceiling
effect, in terms of capacity and SINR, seen previously in limited feedback literature~\cite{Jindal:a,Love:a}.

To ensure the decreasing nature of SMSE, the receivers have to decrease the quantization error
proportionately to the increase of signal power. This condition can be met by increasing feedback bit with varying power
so that $ 2^{-\frac{B}{M-1}} P_{max} $ remains constant. This relation of feedback bits and varying power was at first noticed in terms of sum-rate in~\cite{Jindal:a}.

 \section{Numerical Simulations}

In this section we compare our proposed scheme with the leading feedback schemes in the literature. Since our proposed algorithm
uses channel resources to know the post-processing information of $\mathbf{U}$ \& $\mathbf{P}$,
we use an MMSE receiver to simulate the other existing algorithms. This preserves the fairness
of the comparisons since the performance of the system always improves with an MMSE receiver for mutually unorthogonal
channels~\cite{Boccardi:b}.
 
As mentioned before, our proposed transceiver for MU - MIMO can be readily generalized to MU - MISO system. In Fig~\ref{fig:BER_MISO}, we compare the performance of the proposed algorithm
to the existing precoders in a limited feedback MU - MISO system. 
The proposed algorithm performs better than the MMSE precoder~\cite{Love:a} by using MSIP quantization and convexity of the power allocation problem. The traditional SMSE transceiver, that ignores quantization error, performs
well at lower SNR, but begins to worsen at a SNR of 15dB. Thus the proposed transceiver 
improves over the state-of-the-art in MU MISO limited feedback precoders.

To the best of our knowledge, coordinated beamforming is one of the very few existing linear transceivers that avoid dimensionality constraint in the MU MIMO with multiple data stream scenario. In Fig.~\ref{fig:CoordBF} we compare
the proposed algorithm with coordinated beamforming.  Since coordinated beamforming
~\cite{Heath} implements joint transceiver design, it
performs better than the proposed algorithm with full CSIT. However,
coordinated beamforming needs at least $ \left ( M^2 - 1 \right) $ bits for the
feedback of $ \frac{\mathbf{\hat{H}} \hat{\mathbf{H}}^H} {||\mathbf{\hat{H}}||^2_F} $. We used 15 bit
per data stream in a MU MIMO system with four transmit antennas. 15 bits
per data stream means 1 bit per unique scalar entry of that matrix which is
very low. Due to large quantization error,
the eigen structure of the channel gets mangled at the BS~\cite{Heath:b} which leads to loss of performance.
On the other hand, the quantization error of the fed back vector in the proposed algorithm always remains less than or equal to
 $ 2^{\frac{-B}{M-1}} = 0.03125 $. Thus, the proposed
algorithm performs very close to its full CSIT curve and outperforms coordinated beamforming
~\cite{Heath} with limited feedback. 

In Fig ~\ref{fig:ZFQBC} we compare our proposed scheme with other VQ
combining limited feedback MU MIMO transceivers. Since to the best of our knowledge, existing VQ combining MU MIMO schemes have not dealt with multiple data streams per user, we stick with one data stream per user in
this comparison. The proposed scheme outperforms Boccardi MET~\cite{Boccardi:a} and Jindal QBC~\cite{Jindal:b} due to the use of SMSE precoder, adaptive receive combining and optimal power allocation. Although our algorithm outperforms Boccardi's MESC~\cite{Boccardi:b} upto 20 dB, ~\cite{Boccardi:b} seems to converge at a lower error floor than the proposed algorithm. This happens because of the lack of actual quantization error variance knowledge at the BS in our proposed algorithm. Due to the adaptive quantization policy of the proposed algorithm,
the quantization error variance changes from low to high SNR. since we only quantize the direction of the effective channels, the quantization error norm is not fed back to the BS. Therefore, the proposed SMSE precoder suffers from the lack of error variance knowledge. The quantization error in Boccadi MESC~\cite{Boccardi:b} also changes from low to high SNR but the BS does need this knowledge due to the use of ZF precoder. 

Our proposed transceiver adds to the literature by allowing multiple data streams per user. Fig~\ref{fig:CombiningTechniques} shows the comparison of our transceiver's performance to other possible methods that can be implemented to transmit multiple data streams per user. In Fig~\ref{fig:CombiningTechniques}, Eigen Based Combining projects the MIMO channel to its dominant eigenvectors to create the effective MISO channels~\cite{Nazmul} and QBC choses the set of codevectors that will generate least amount of quantization error as effective MISO channels~\cite{Jindal:b}. The proposed transceiver approaches Eigen Based Combining at low SNR and QBC at high SNR. Thus the proposed algorithm retains the advantages of both Eigen Based Combining and QBC by providing a trade-off between signal power, intra and inter user interference.

\section{Conclusion}

In this paper, we proposed linear transceiver design in the downlink of a MU MISO and MU MIMO system (with multiple data streams per user) using SMSE precoder at the BS, MSIP VQ as the feedback algorithm and MMSE decoder at the receivers.
However, to encode the channel information, the receivers use MESC first. In the MU MISO, the individual users send back the indexes of their quantized channels to the BS. In the MU MIMO scenario, the users convert their MIMO channels to effective vector downlink MISO channels to maximize the expected SINR and then send the indexes of these quantized MISO channels. The BS uses the quantization error of MSIP in the SMSE precoder design and finds the downlink precoder and power allocation vector using a convex optimization problem. The proposed designed system was shown to outperform the previously existing linear transceivers in the MU scenario for limited feedback, while also allowing for multiple data streams per user.

One possible extention of the present work will be the detailed analysis of the expected quantization error variance in the 
intermediate SNR's. The way the receivers find the trade-off between signal power increase, and intra and inter user interferrence reduction will give an insight to analyze this problem.

In this work, only shape feedback is sent to the BS. This is reasonable if the average channel magnitudes of all user is equal. The reason lies in the fact that the alignment of precoding vector with channel direction is more important than the power allocation. However, this assumption may remain valid only in a small scale fading scenario. In a more practical scenario, different users will be located at different distances from the BS and therefore average channel magnitude will be different. In that case, bit allocation in the channel magnitude will also be important. An extension of the present work will be the optimization of feedback bits among the channel magnitude and direction information.

Our present work can also be extended to a slowly time varying channel. The temporal correlation between different blocks can be used to reduce the amount of feedback overhead.

\newpage

\bibliographystyle{IEEEbib}
\bibliography{bib_overall}

\section{Figure Captions}

1. Block Diagram of Multiuser MIMO Downlink

2. Block Diagram of Multiuser MIMO Uplink with channel and decoder combined as a whole block

3. SMSE analysis of the proposed precoder, $ M = 5$, $K = 5$, $N_k = 1$, $L_k = 1$ $\forall k$, $B = 10$, QPSK

4. Comparison with previous MU-MISO precoding techniques with $M = 4$, $K = 4$, $L_k = 1$ $\forall k$, $B = 10$, QPSK

5. Comparison with the coordinated beamforming $ M = 4$, $K = 2$, $N_k = 4$, $L_k = 1$ $\forall k$, $B = 15$, QPSK

6. Comparison with previous MU-MIMO VQ precoding techniques $ M = 4$, $N_1~=~N_2~=~2$, $N_3~=3$, $L_k~=1$, $\forall k$, $B = 15$, QPSK

7. Different receive combining techniques with multiple data streams per user $ M = 4$, $N_k = 3$, $L_k = 2$, $\forall k$, $B = 12$, BPSK

\newpage

\begin{figure}[t]
 \epsfig{figure=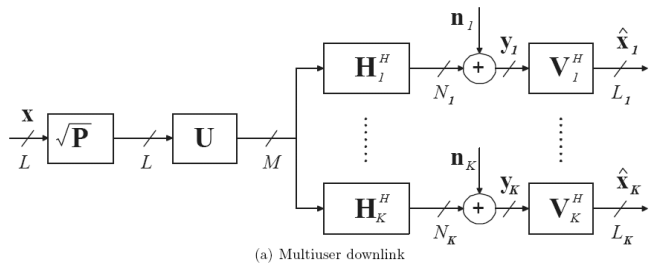,width=85mm}
 \caption{Block Diagram of Multiuser MIMO Downlink}  \label{fig:MUdownlink}
\end{figure} %
\begin{figure}[t]
 \epsfig{figure=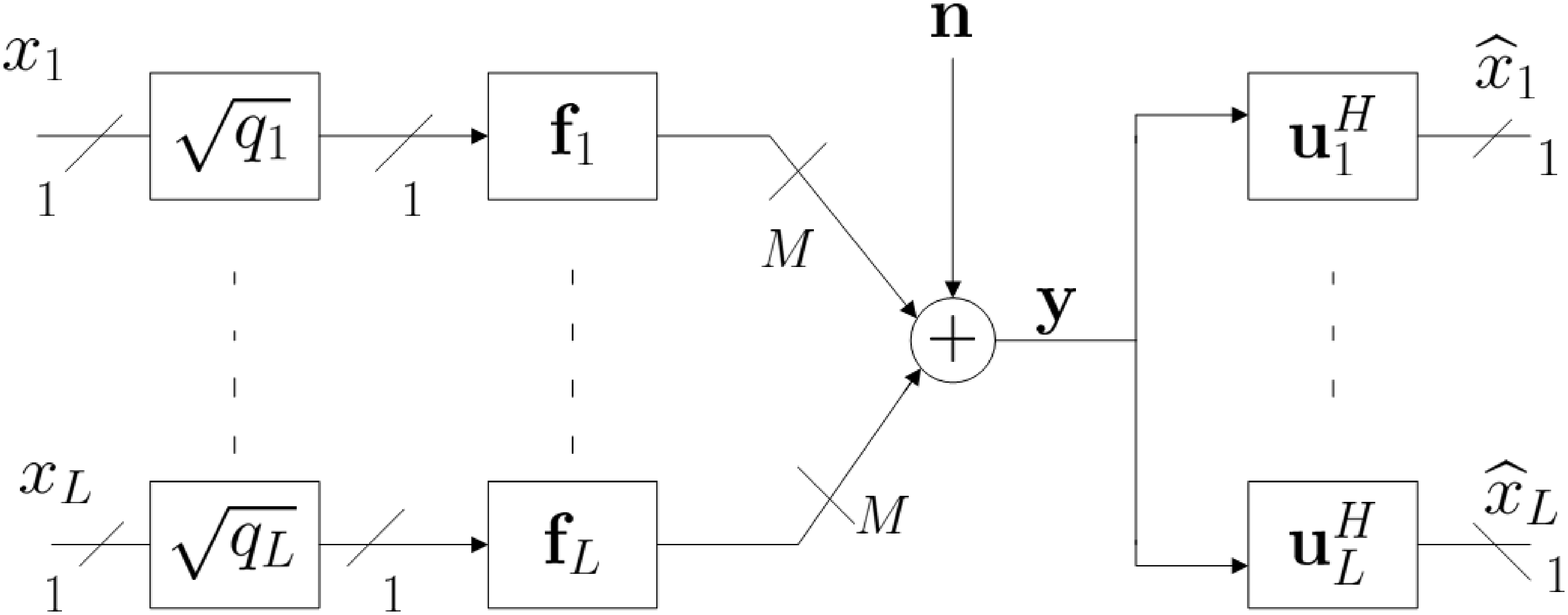,width=90mm}
 \caption{Block Diagram of Multiuser MIMO Uplink with channel and decoder combined as a whole block}  \label{fig:MUuplink}
\end{figure} %

\begin{figure}[t]
\epsfig{figure=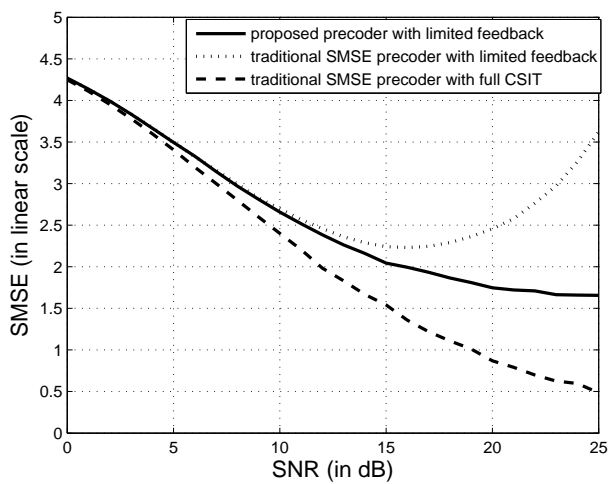,width=90mm}
\caption{SMSE analysis of the proposed precoder, $ M = 5$, $K = 5$, $N_k = 1$, $L_k = 1$ $\forall k$, $B = 10$, QPSK}   \label{fig:SMSE_analysis}
\end{figure}

\begin{figure}[t]
 \epsfig{figure=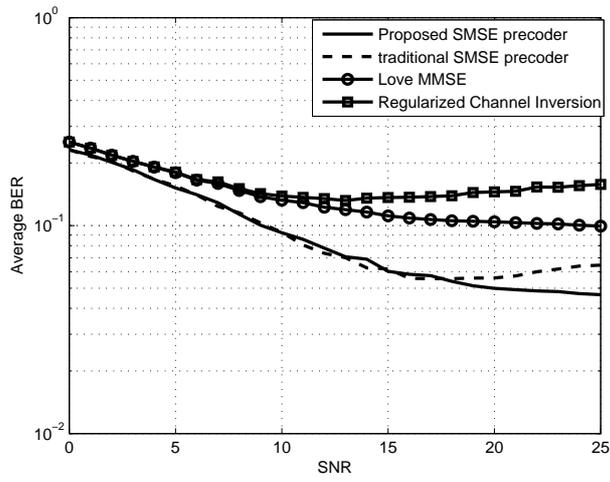,width=90mm}
 \caption{Comparison with previous MU-MISO precoding techniques with $M = 4$, $K = 4$, $L_k = 1$ $\forall k$, $B = 10$, QPSK}
 \label{fig:BER_MISO}
 \end{figure}
 
 \begin{figure}[t]
 \epsfig{figure=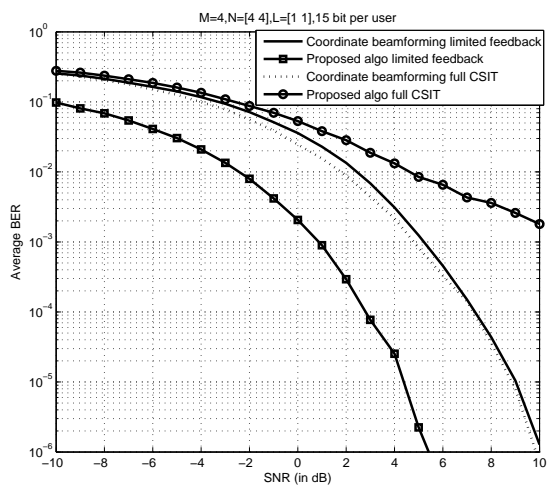,width=82mm}
 \caption{Comparison with the coordinated beamforming $ M = 4$, $K = 2$, $N_k = 4$, $L_k = 1$ $\forall k$, $B = 15$, QPSK} \label{fig:CoordBF}
 \end{figure} %
 
 \begin{figure}[t]
\epsfig{figure=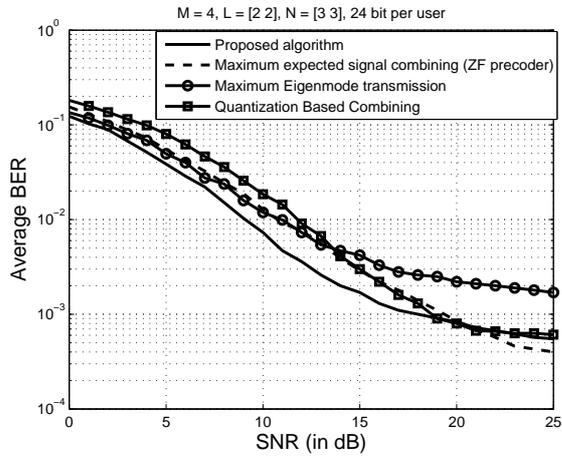,width=82mm}
\caption{Comparison with previous MU-MIMO VQ precoding techniques $ M = 4$, $N_1~=~N_2~=~2$, $N_3~=3$, $L_k~=1$, $\forall k$, $B = 15$, QPSK} \label{fig:ZFQBC}
\end{figure} %

\begin{figure}[t]
 \epsfig{figure=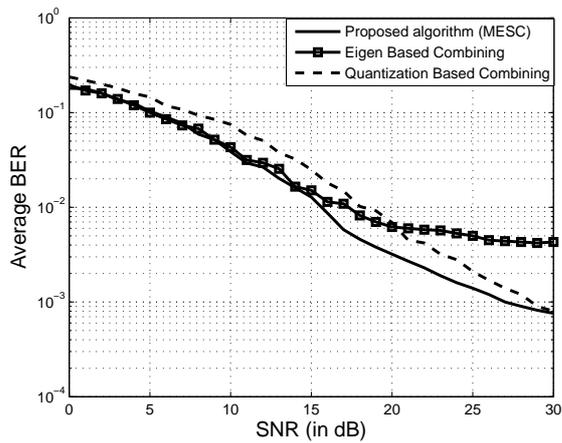,width=82mm}
 \caption{Different receive combining techniques with multiple data streams per user $ M = 4$, $N_k = 3$, $L_k = 2$, $\forall k$, $B = 12$, BPSK} \label{fig:CombiningTechniques}
\end{figure}

\end{document}